# Application of Model Derived Charge Transfer Inefficiency Corrections to STIS Photometric CCD Data

P. Bristow

July 2003

## Abstract

*The STECF Calibration Enhancement effort for the Space Telescope Imaging Spectrograph (STIS) aims to improve data calibration via the application of physical modelling techniques. As part of this effort we have developed a model of the STIS CCD readout process. The model itself is described in some detail in earlier ISRs. Here we consider the application to STIS photometric data. We demonstrate that this approach can successfully remove the trails typically seen in CTE affected data and go on to ask whether this charge is being restored to the cores of detected objects in the correct quantities. We find good agreement between the simulation derived corrections and empirical corrections for a range of background and source levels. Moreover we find that, where the two differ, it is because only the simulation properly accounts for the charge distribution in the array.*

# 1. Introduction

Charge Coupled Devices (CCDs) operating in hostile radiation environments suffer a gradual decline in their Charge Transfer Efficiency (CTE, or an increase in charge transfer inefficiency, CTI). STIS and WFPC2 on HST have both had their CTE monitored during their operation in orbit and both indeed show a measurable decline in CTE which has reached a level which can significantly affect scientific results (e.g. Cawley et al 2001, Heyer 2001, Kimble, Goudfrooij and Gililand 2000).

Instead of seeking empirical corrections we have chosen to construct a physical model which allows us to predict the effects of poor CTE in astrophysical data using STIS as a test case. Detailed discussion of the model development and the physics involved can be found in Bristow & Alexov et al. 2002 and Bristow 2003a. Here we restrict the discussion to a review of the performance of the model when used to correct point sources in STIS CCD imaging data, particularly in comparison to the empirical algorithmic correction. (Bristow 2003c presents a similar discussion regarding STIS CCD spectroscopic data).

There has been a major effort by the WFPC2 and STIS groups at STScI to monitor and characterise the CTI effects seen in the data from these instruments (e.g. Ferguson 1996, Whitmore 1998, Whitmore et al. 1999, Kimble, Goudfrooij and Gililand 2000). This has resulted in well calibrated empirical corrections for CTI affected data (Dolphin 2000, Dolphin 2002 and Goudfrooij and Kimble 2002, hereafter GK2002).

GK2002 present a detailed analysis of CTI effects in STIS data specifically obtained for this purpose and derive an empirical algorithm for correcting CTE loss in photometry of point sources. As discussed in Bristow & Alexov (2002) and Bristow (2003a), empirical corrections for STIS provide both a source of fine calibration and a benchmark for the performance of the rather more expensive (in terms of CPU time and user interaction) model based corrections.

Empirical corrections give the fractional CTE loss as a function of signal strength (for a point source), background, epoch and position on the chip (distance from the readout register. It is necessary to formulate and calibrate the corrections differently for photometric and spectroscopic data because of the differing nature of the flux (and therefore charge) distributions. Moreover, empirical corrections only apply to point sources. By modelling the readout process we are able to correct for any charge distribution and therefore can apply this method to all data whether photometric or spectroscopic.

However, we should expect that the model based corrections are in general agreement with the empirical corrections for point sources. Indeed, by demanding such a general agreement we can use the empirical corrections to calibrate the physical model. This is much easier than returning to, and re-analysing, the data itself as the empirical corrections are essentially a distillation of what is to be learnt from the data with respect to CTI. That is not to say that the physical model is simply constrained to reproduce the empirical corrections. However our hypothesis is that if the CTE model reproduces empirical results on average for point sources, then it is reasonable to conclude that it is correctly modifying the charge distribution and will also therefore correctly predict the CTI in extended sources and indeed the whole image array.

Instances of disagreement between model and empirical results are interesting as they will tell us if the model based correction is either failing (if we can see no good reason for the disagreement) or improving upon the empirical corrections by predicting differences in the charge distribution which empirical corrections could not have dealt with.

In this paper we apply our simulation derived corrections to STIS calibration data which was obtained for the purpose of monitoring CTE. In section 2 we show typical cosmetic damage caused by poor CTE and present the results of cleaning these images with our corrections. A quantitative analysis follows in section 3 where we compare the empirical and simulation derived corrections for extracted sources. Section 4 considers discrepancies between empirical and simulation derived corrections and identifies the causes.

## 2.Cleaning CTE trails

Figure 1a is an example of a section of a STIS image array displaying clearly visible CTI trails. The trails are caused by the loss of electrons from the large charge packet representing a source as it is transferred (in this case) up the chip to the readout register. The charge lost is held in traps in the silicon lattice along the entire column between where the charge was collected and the register. When the smaller charge packets traverse these pixels there is a chance that some of the trapped charge will be emitted. This process is described in somewhat more detail in Bristow & Alexov et al. 2002 and Bristow 2003a. Simulation of this process for a given charge distribution is essentially the only way to remove this kind of cosmetic CTI effect, there is no simple way to characterize it for empirical fits to be possible.

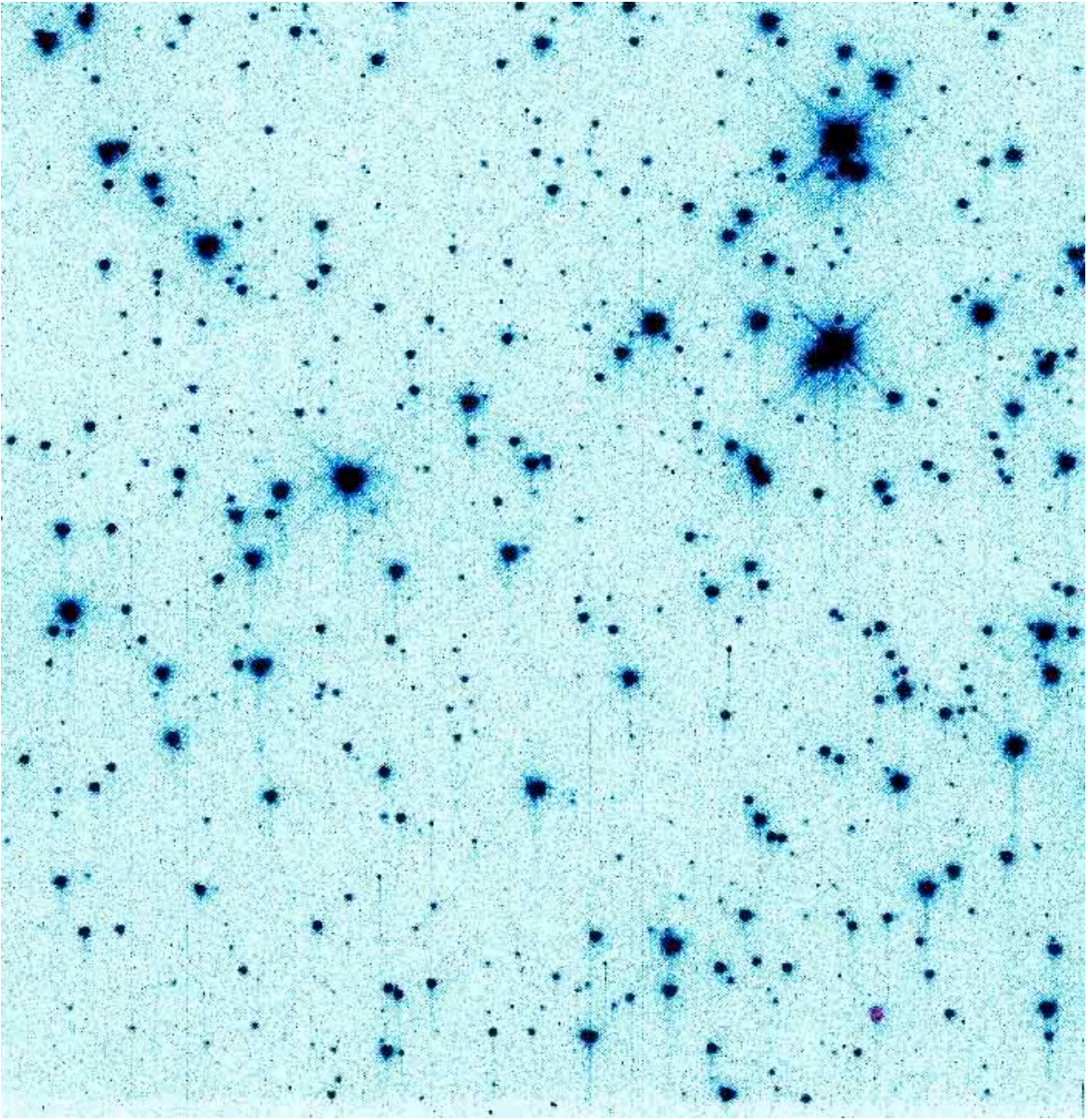

• Figure 1a: Section of CTI affected STIS CCD imaging mode data

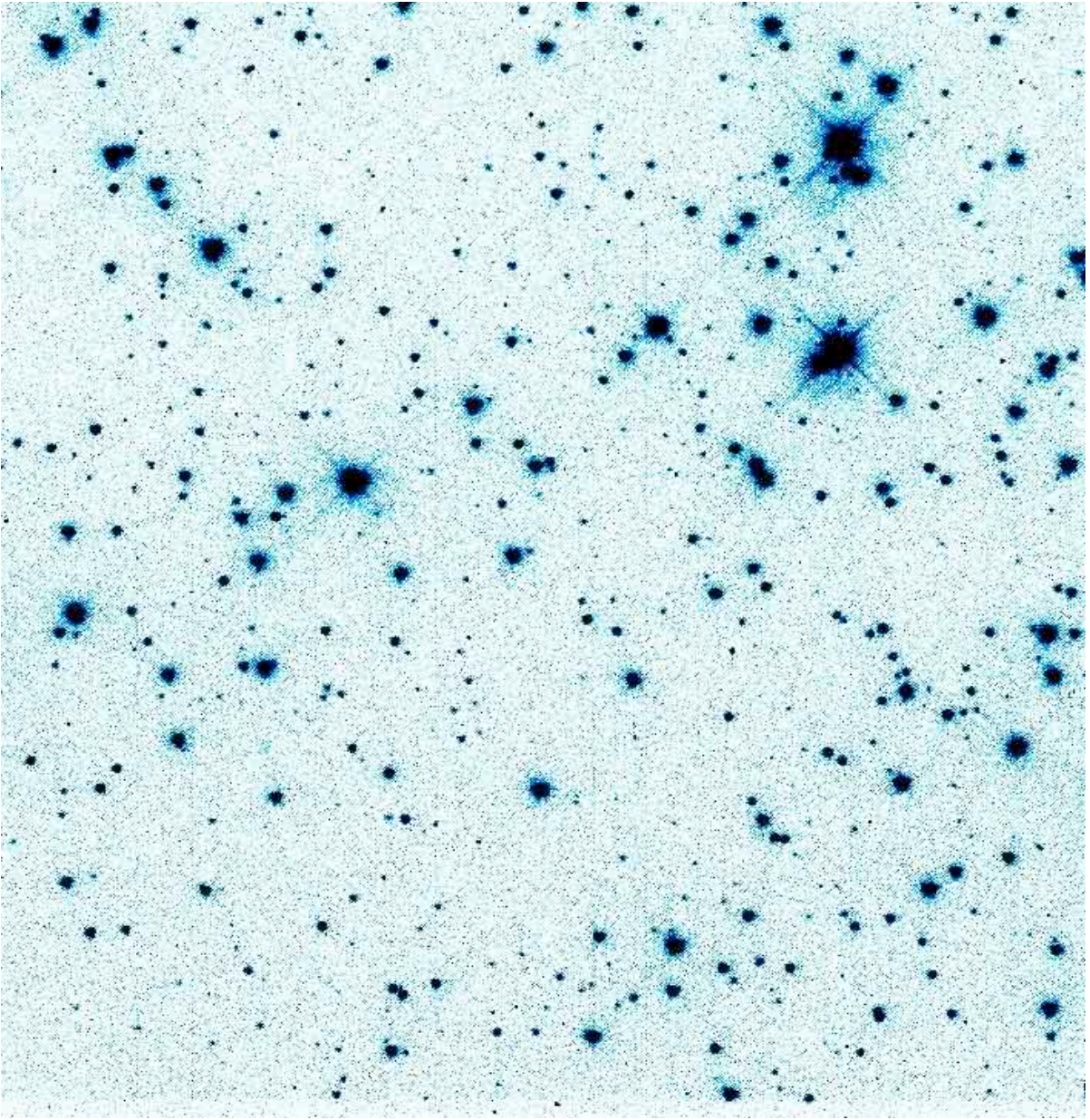

· Figure 1b Image data section of Figure 1a after application of simulation derived correction

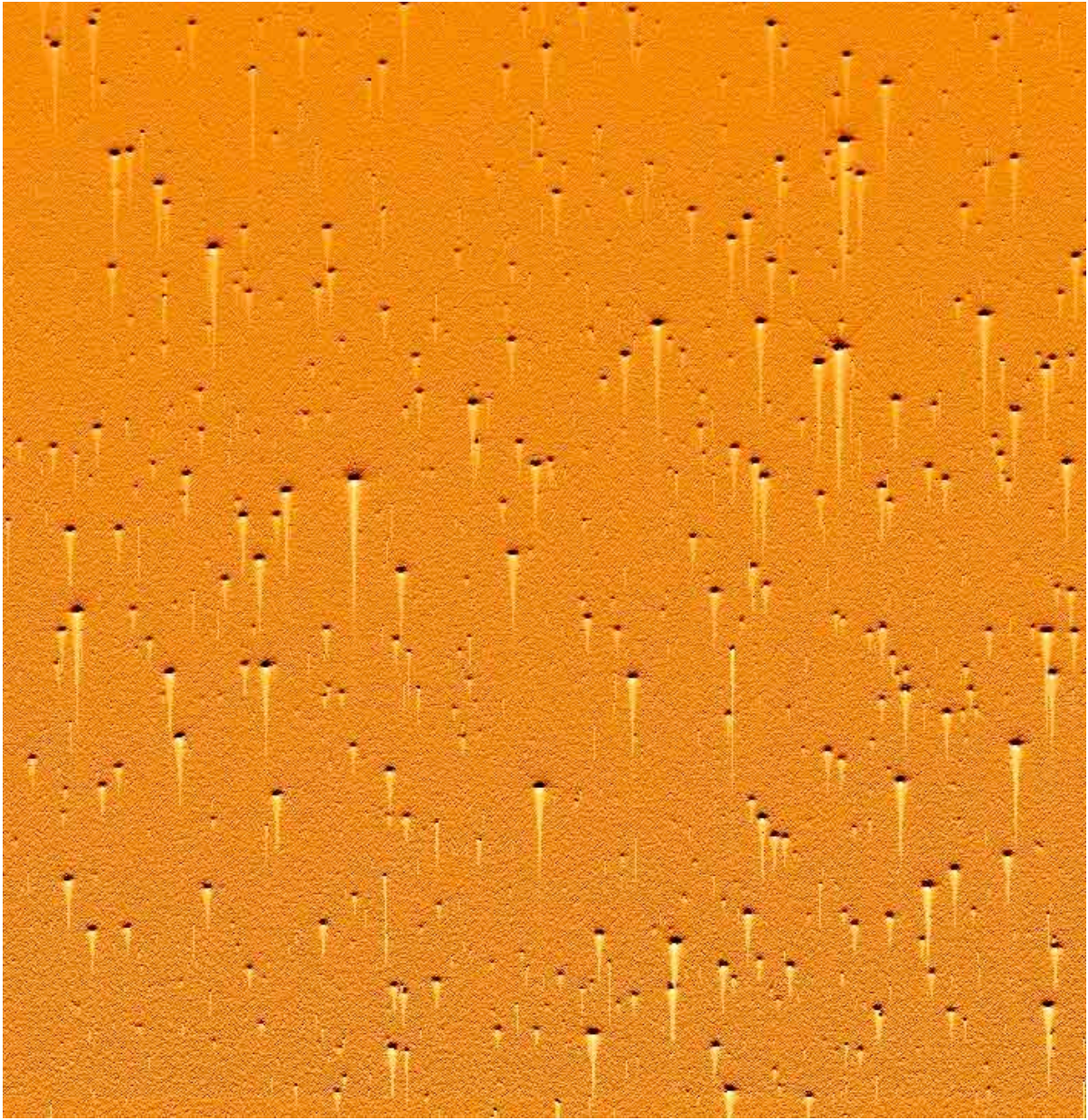

Figure 1c Difference Image for Figures 1a and 1b

The results of the simulation derived correction upon the image array are presented in figure 1b. The trails are very much reduced, though still visible in some cases. We can tune the model simply by altering the density of traps (silicon defects) assumed (or by adjusting any of a number of other parameters which affect the readout process in less obvious ways, see Bristow 2003a). Increasing the trap density will indeed remove the trails more completely; on the other hand, increasing too much will invert the trails so that a shadow seems to trail sources. From the modelling point of view this is the same as saying, increasing the trap density results in larger trails in the simulation prediction, the correction deduced from this then includes a larger amount of trail removal. However, rather than setting the trap density at the level which appears to produce the best "trail free" images, we use the more quantifiable agreement between photometry from empirically corrected sources and those detected in simulation corrected images to fine tune model parameters. Bristow 2003a describes how the best model parameters were ascertained from a number of considerations, including the tests performed below. Figure 1b was derived with this best set of parameters. Figure 1c is simply a difference image of 1a and 1b (1c=1a-1b), here we see clearly the dark area on the leading edge of the source (nearest to the readout register, at the top in this image) where the readout process causes the loss of signal and the bright trail where charge is deferred during readout.

A further point to note here is that hot pixels and cosmic rays in figure 1a can also be seen to have CTI trails, in fact close inspection of raw data reveals that even noise spikes have trails. Whilst hot pixels and cosmic rays are removed as part of the standard CALSTIS pipeline reduction, their trails are not. If one knows where to look it is possible to see these trails in the calibrated data products. This has implications for the noise properties (see Bristow 2003a for a further discussion of CTI noise).

The cleaned image in figure 1b gives us qualitative confirmation that the simulation derived correction can clean up CTI trails. In order to get a quantitative verification we ask whether it puts the trail charge back into the object core in the correct quantity.

# 3. Comparison with Empirical Corrections

In order to see how successfully the simulation derived corrections restore the photometric properties of CTI degraded sources we use the empirical corrections of GK2002 as a comparison. These corrections have been calibrated for a range of epochs, background levels, source levels and distances from the readout register for a large number of datasets by using the ability of the STIS CCD to readout in two directions to registers at the top or bottom of the chip. As such they represent the average level of CTI effects for many datasets, saving us the trouble of having to extract this information from the data ourselves.

## 3.1 Source Detection and Application of Detections

Comparison for a large number of sources in several datasets was facilitated by extending the pipeline scripts which run the simulation on the charge distribution of a given dataset (see Bristow 2003a) to include the following post processing steps:

1. SExtractor is run over both the raw and simulation corrected images. Selection is *all* objects with a flux (7 pixel aperture) more than 4.0_ above the background. There is no size limit, at this stage we want to include all objects, even hot pixels. This allows us to see in some cases where a hot pixel with a CTI trail and therefore >1 pixel) in the uncorrected image matches with a single pixel in the corrected image.

2. The two extracted source lists are x-correlated:

   - Centroid y-position within 1 pixel

   - Centroid x-position within 0.5 pixels

3. The ratio of flux (7 pixel aperture and 'best') between corrected and uncorrected images for matched sources is computed.

4. Empirical correction (ratio corrected to uncorrected) is calculated using the epoch, y-position, flux and background (local and global).

5. A combined table is produced with the following entries:

| Number | Parameter |
|---|---|
| 1 | Object number |
| 2 | X position in raw data (pix from left) |
| 3 | Y position in raw data (pix from bottom) |
| 4 | Aperture flux in raw data (ADU) |
| 5 | Peak flux in raw data (ADU) |
| 6 | Area within 4 _ isophote in raw data (sqr pix) |
| 7 | Aperture flux in raw/Aperture flux in corrected data |
| 8 | Aperture flux/empirically corrected flux computed with global sky background |
| 9 | Aperture flux/empirically corrected flux computed with local sky background |
| 10 | 7/8 |
| 11 | 7/9 |
| 12 | SExtractor "best" flux in raw data/SExtractor "best" flux in corrected data |
| 13 | 12/8 |
| 14 | Peak flux in raw data/peak flux in corrected data |
| 15 | FWHM in raw data - FWHM in corrected data(pix) |
| 16 | Elongation in raw data / elongation in corrected data |
| 17 | Isophotal area (4_) in raw/Isophotal area (4_) in corrected |
| 18 | X position in raw data - X position in corrected data |
| 19 | Y position in raw data - Y position in corrected data |
| 20 | Star-galaxy classification in raw data |
| 21 | Star-galaxy classification in corrected data |
| 22 | Local background in raw data |
| 23 | Local background in corrected data |
| 24 | 22/23 |
| 25 | Corrected data extraction list no. |
| 26 | Raw data extraction list no. |

6. A linear least squares fit is made to empirical correction (local background) versus simulation correction (7 pixel aperture) for all sources satisfying the following criteria:

   - Area above threshold >= 4 pixels
   - Peak flux > 100 ADU
   - Star-galaxy classification > 0.75

For each dataset the fit parameters are recorded along with the mean local background and number of sources fulfilling the criteria.

In the above, the star-galaxy classification (hereinafter SGC) and "best" flux estimates are those offered by the SExtractor package, see Bertin (??) for further details. We compute the empirical corrections for both the local and global estimates of the background level from SExtractor.

For each dataset the fit parameters are recorded along with the mean local background and number of sources fulfilling the criteria.

We have used the multi-parameter output of step 5 to gain an insight into the interdependencies between the derived and measured parameters. Some of the more useful plots are presented and discussed in section 3.3 below.

## 3.2 Results

In figure 2 we present a comparison of the empirical and simulation derived corrections for all sources satisfying the criteria of table 1, detected in the datasets summarised in table 2. The dashed line represents perfect agreement.

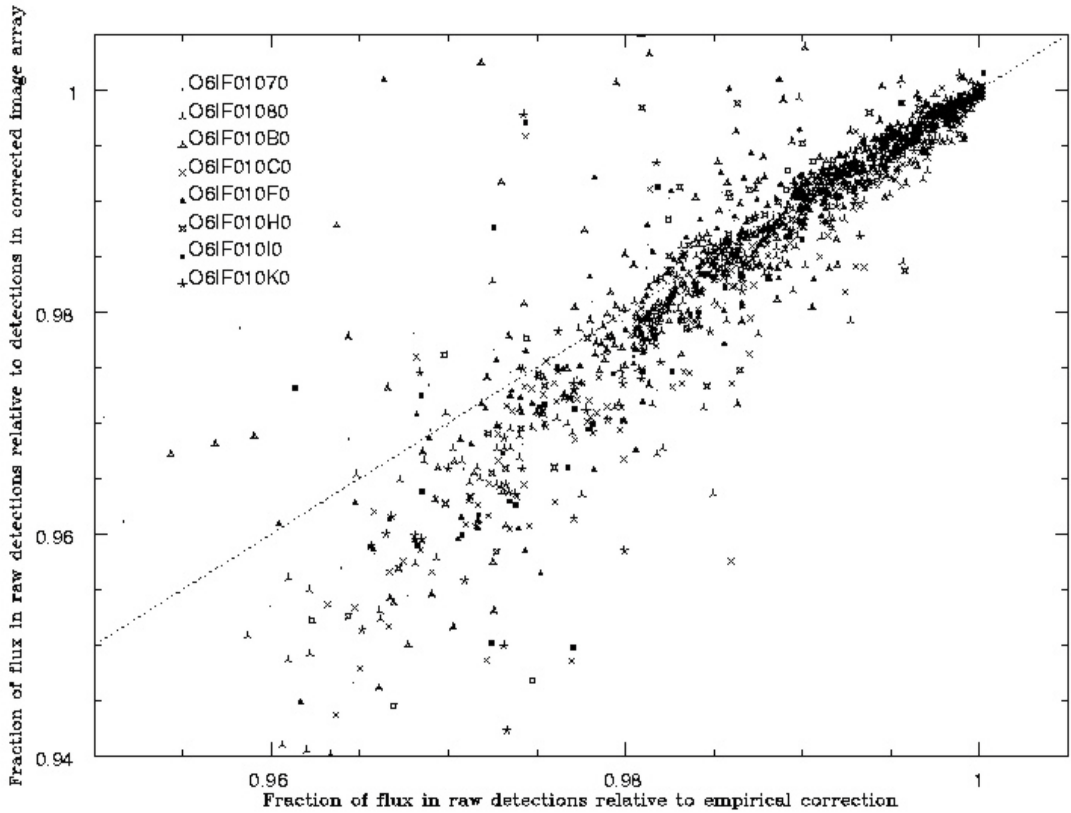

Figure 2a Comparison between model and empirical photometric corrections for the flux within a 7 pixel diameter aperture. Detections come from the datasets according to the key in the figure. The dotted line represents perfect aggreement between model and empirical corrections. Selection criteria are those of table 1.

• Table 1: selection limits

| Parameter | Limit |
|---|---|
| Detection threshold | 4 |
| Area(pixels above detection threshold) | >4 |
| SGC in corrected dataset | >0.9 |
| Minimum peak flux (ADU above background) | >100.0 |

• Table 2: Datasets

| Dataset   | Readout Amp | Exposure Time (s) |
|-----------|-------------|-------------------|
| O6IF01030 | D           | 495.0             |
| O6IF01070 | D           | 20.0              |
| O6IF01080 | D           | 100.0             |
| O6IF010B0 | B           | 20.0              |
| O6IF010C0 | B           | 100.0             |
| O6IF010F0 | B           | 100.0             |
| O6IF010H0 | B           | 20.0              |
| O6IF010I0 | B           | 100.0             |
| O6IF010K0 | B           | 20.0              |

On the whole the points in figure 2 show a good agreement between empirical and simulation derived corrections. There is however some scatter and a systematic overestimate from the simulation derived corrections relative to the empirical corrections as the corrections become larger (lower values in the plot). The scatter merely reflects the fact that the simulation deals with the actual, inherently noisy, charge distribution. In section 4 below, we look at some of the more extreme examples and trace their cause back to anomalies in the charge distribution of the raw data.

Regarding the systematic disagreement, it is worth pointing out that the level of agreement between the empirical and simulation derived corrections depends to some extent upon the selection of sources. Restricting the selection to larger, brighter objects with higher SGC always results in a closer agreement as seen in figure 3 (peak flux>1000ADU/pixel, area>40pixel, SGC>0.95). This is to be expected if the empirical corrections were calibrated mainly using relatively bright stars. For this reason we do not consider the apparent discrepancy seen for lower values in figure 2 to be significant.

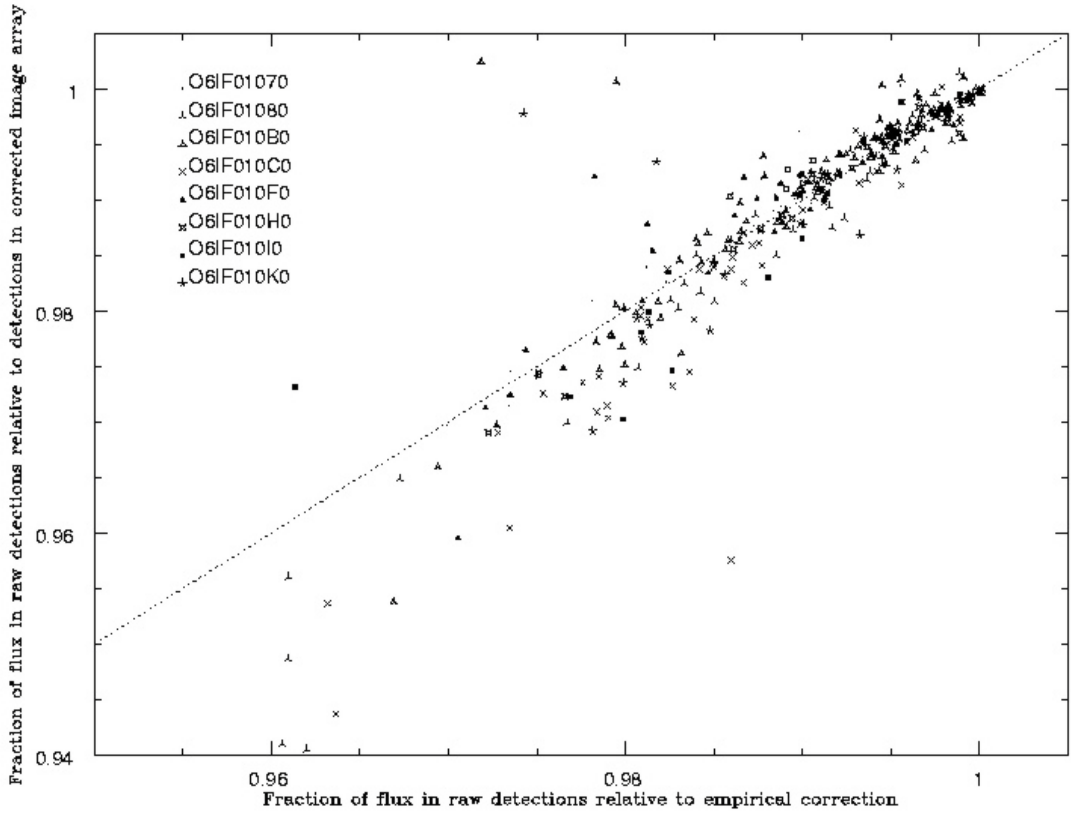

Figure 3 As figure 2, except that only sources with a high star-galaxy classification are plotted.

## *3.3 Multivariable plots*

We can learn a lot more from the results produced by the process described in section 3.1. In what follows, the criteria of table 1 were applied except where otherwise stated. It is useful to define here:

$$R = \frac{\text{Flux in source measured from simulation corrected image array}}{\text{Flux predicted by empirical correction}}$$

Clearly $R=1$ for sources which the simulation corrected image array is in perfect agreement with the empirical correction.

Figures 4a-d shows the ratio $R$ as a function of the parameters which are important to the calibration of the empirical corrections. In 4a this ratio is plotted against row number (y), reassuringly there is little dependence in the mean value (close to 1.0) of this ratio upon row number; however the scatter increases as the distance from the readout amplifier increases.

Figure 4b presents the dependency upon local background (as measured by SExtractor in an annulus surrounding the source of thickness 24 pixels). Here we can see the sources from datasets with different exposure times separate out into groups. Within each group there is then a smaller range of background level. The longer exposure groups show little dependence of this ratio (again close to 1.0) upon background, though within the groups there is more scatter at lower values of local background. The shortest exposure group has a mean ratio slightly below 1.0; this reflects the discrepancy seen in figure 2.

Figure 4c shows the dependency upon source level. Once again there is little variation in the mean, but more scatter at lower levels of source level. Finally figure 4d presents the dependence upon SGC. For this plot we have obviously removed the restriction upon star-galaxy classification for selection and have also reduced the minimum peak flux to 10ADU. The mean ratio seems to drop slightly below 1.0 for lower values of SGC. This is however to be expected as the empirical corrections were calibrated for stellar like sources.

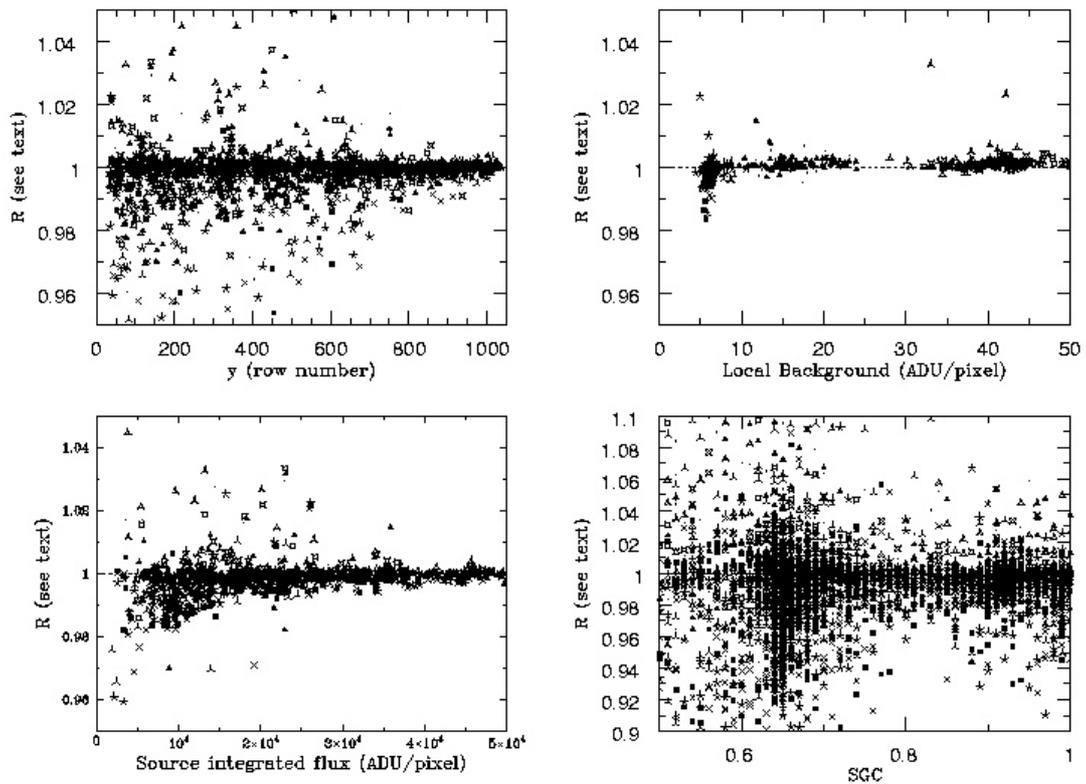

- Figure 4 a: R (see text) versus row number. b: R versus local background. c: R versus source flux. R versus star-galaxy classification. Selection criteria are those of table 1 for a-c. For d there is no limit on star-galaxy classification.

Figures 5 and 6 verify expected properties of the simulation derived corrections. In 5 the ratio of peak flux in the raw data to that in the simulation corrected data is plotted as a function of row number. The expected dependence is clearly seen. In 6a the y-offset between the raw data and the simulation corrected data is plotted as a function of row number. Once again the dependence is clearly seen whilst the equivalent plot for x-offset in 6b shows no such dependence.

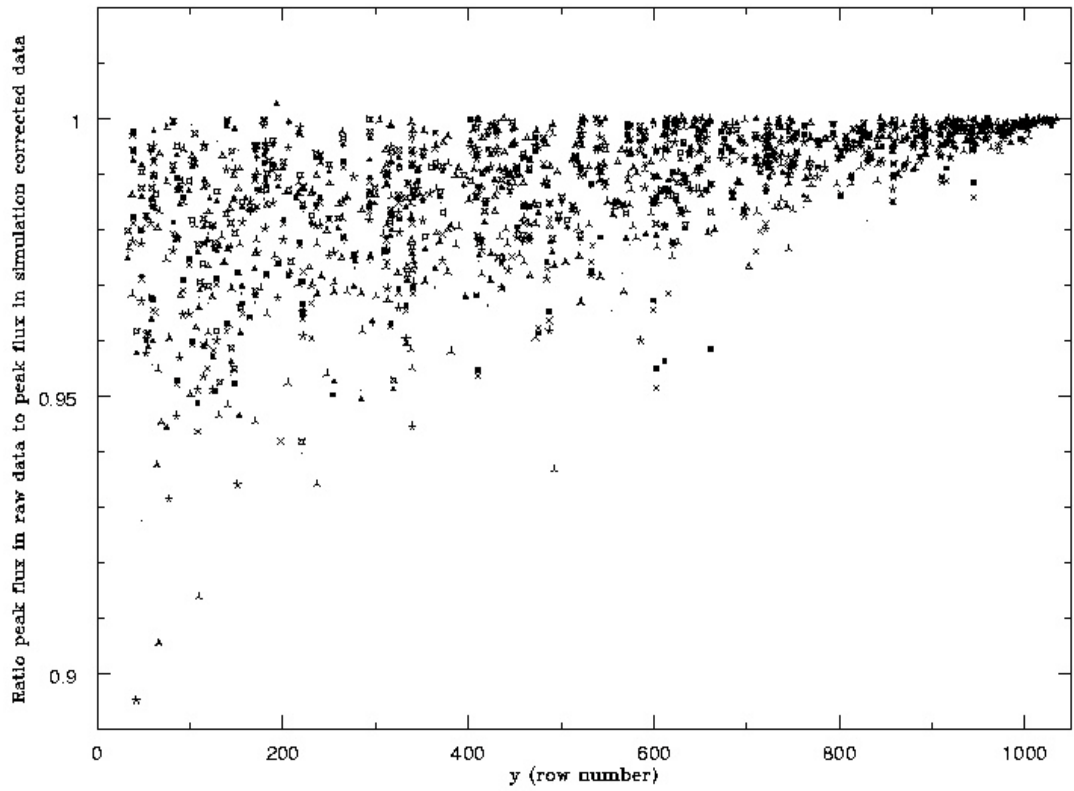

• Figure 5 Simulation derived correction to peak flux versus row number

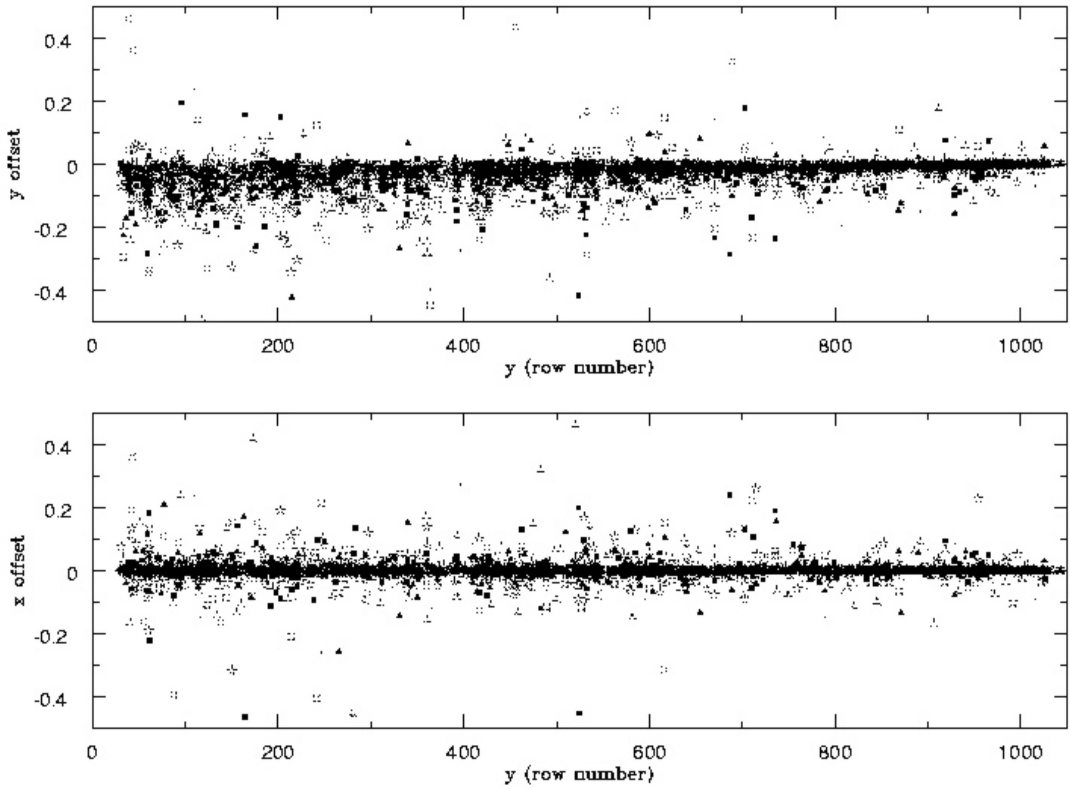

Figure 6 a: Y (row number) offset between centroid of source detected in raw data and its match in the corrected data versus row number. b: X (column number) offset between centroid of source detected in raw data and its match in the corrected data versus row number.

## 3.4 Direct comparison to CTE measured from B/D amplifier pairs

The ability of STIS to read out in two directions to either of two registers, one at the top of the CCD array, connected to the C and D amplifiers, and one at the bottom, connected to the A and B amplifiers, provides a convenient way of monitoring CTE. Indeed this was central to the monitoring of STIS (Kimble et al 2000) and the subsequent derivation of empirical corrections in GK2002. For one pair of datasets (identical field of view and exposure time, one, O6IF01080, B amplifier, the other, O6IF010C0, D amplifier) we follow Kimble et al and plot in figure 7a the ratio of D amplifier to B amplifier flux as a function of row number for each source. As expected there is a clear dependency on row number. Sources near the bottom of the chip have lost considerable charge during the long transfer process in the D amplifier data, but relatively little in the B amplifier data and vice versa. Figure 7b is the same as 7b except now the fluxes are measured in the simulation corrected data. Now the dependency has nearly vanished, but there is still plenty of scatter due to the shot noise in each source.

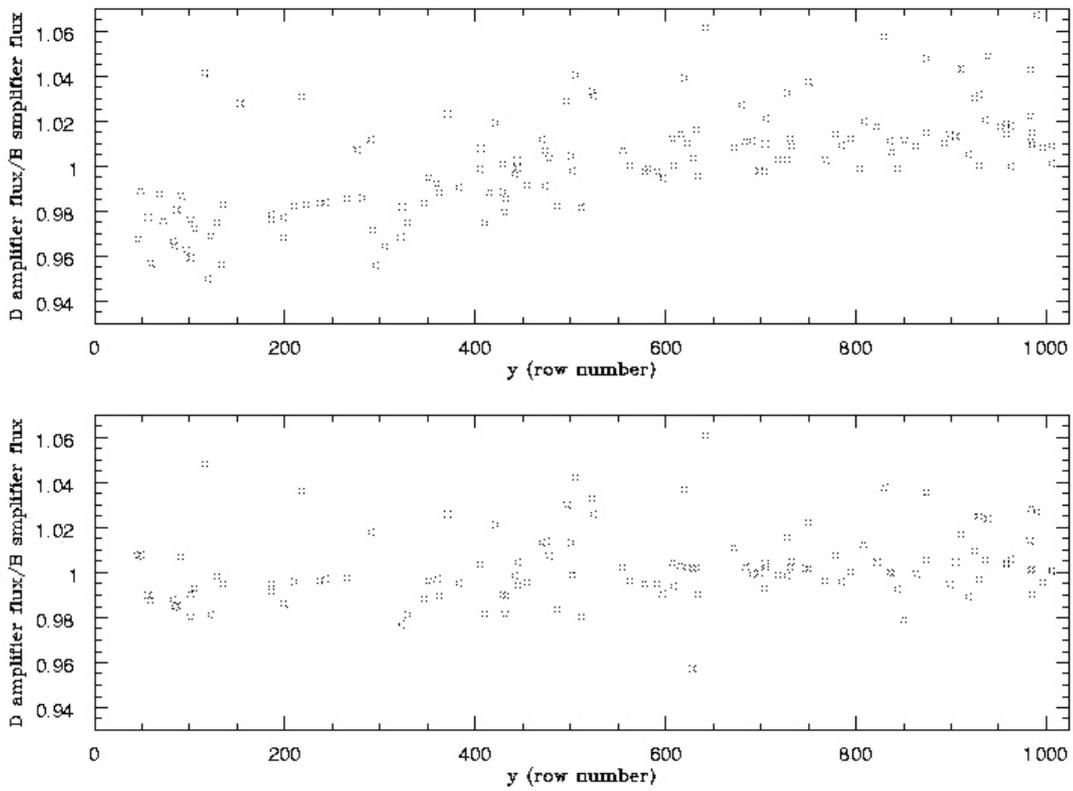

Figure 7 a: Ratio of raw D amp data flu to raw B amp data flux versus row number. b: Ratio of simulation corrected D amp flux to simulation corrected B amp flux versus row number

We can take further advantage of the bidirectional readout and consider just the bottom 100 rows for which B amp data can be considered, to a good approximation, CTI effect free. Taking the ratio of D amplifier flux to B amplifier flux for sources in this part of the image array should give us the CTE suffered by each of those sources. However, in figure 8 where we have plotted these values against the empirical CTE (as crosses) we see that there is considerable scatter. Shot noise differences between the two exposures make it very difficult to make such a direct comparison. Also in this plot are points (triangles) where we have calculated the ratio using the simulation corrected version of the D amplifier data. We might hope that all of these points would have a ratio value of 1.0. Whilst they have moved in that direction, once again, the shot noise difference between the two exposures dominates.

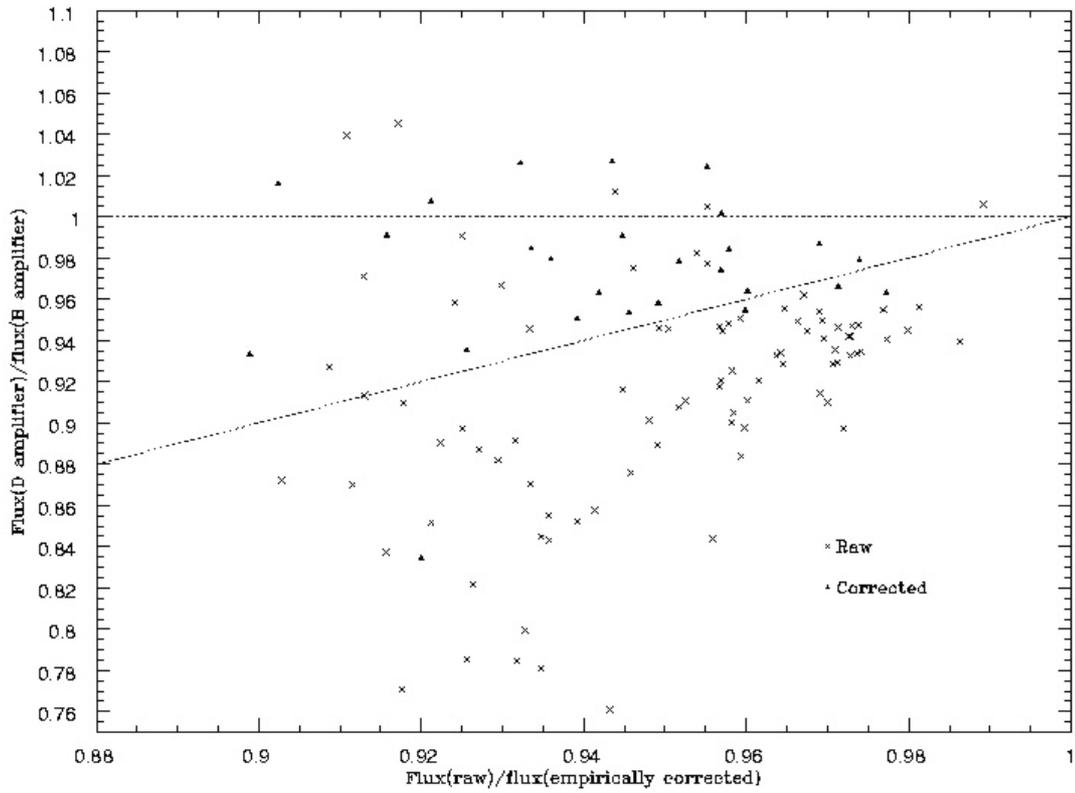

.

Figure 8 Ratio of D amp flux to Raw B amp flux for sources within 100 rows of the B amp register. The Damp flux is that measured in the raw for crosses and that measured in the simulation corrected data for triangles.

# 4.Effects of the charge distribution upon CTE corrections

As noted above, there is considerable scatter in figure 2. This derives from the fact that the non uniformity of the charge distribution causes the CTI experienced by each source to vary in a way which cannot be accounted for in the empirical corrections.

In order to illustrate this we have labelled some of the outlying sources in figure 9a, which is the same as figure 2 except that it only contains data from O6IF01070 to reduce the confusion. The sources labelled α and β have simulation derived corrections which is somewhat smaller than those which the empirical algorithm would assign to them. We can see why by looking at the raw image data from which they were extracted in figure 9b. The sources are highlighted with green boxes, α is the lower one and β is above and to the right. Both are situated just below a larger source. In the readout process, the source above (and nearer to the D readout amp in use) leaves charge behind which reduces the charge loss of sources a and b. In figure 9c we see the two sources in the difference image (as figure 1c above), i.e. the charge lost (dark) and gained (light) during the readout process according to the simulation. Both sources can be seen to lie in the CTE trail of the source above.

Figure 9a: detail of Figure 2 with some of the outlying points highlighted.

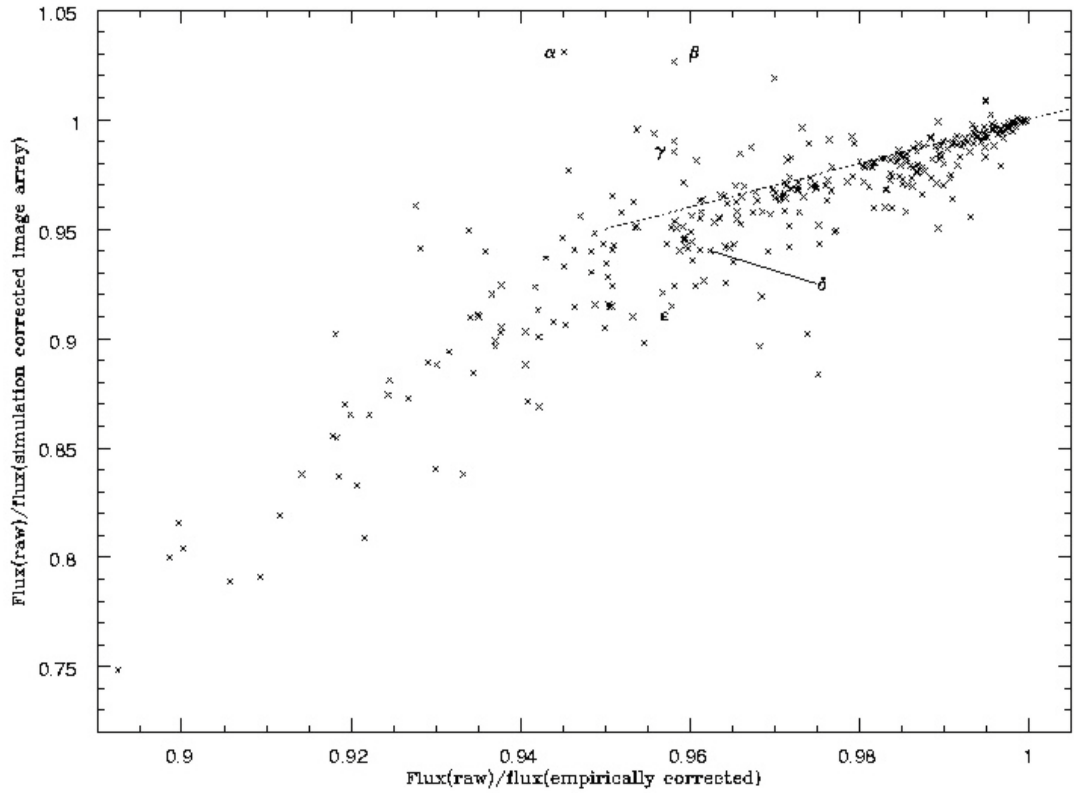

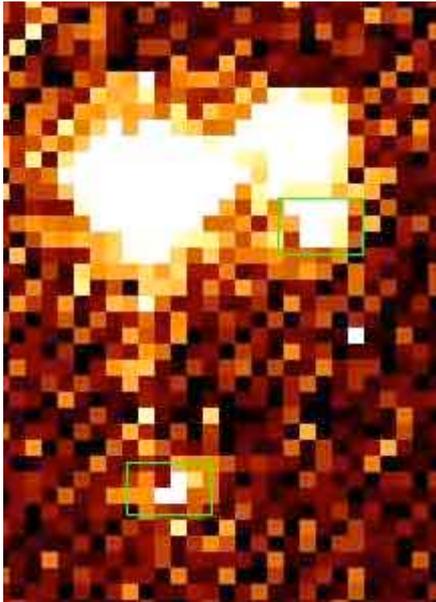

Figure 9b: Sources corresponding to points α (lower left) and β (upper right), in 9a as they appear in the raw data.

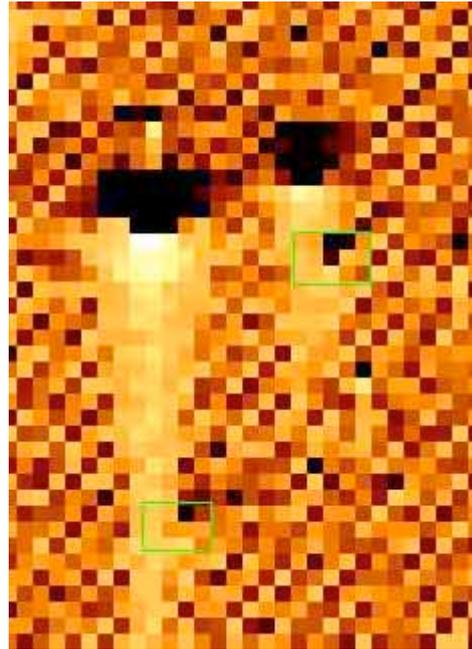

Figure 9c: 7 pixel tall extraction corresponding to points α and β, in 9a as they appear in the difference image generated by the simulation.

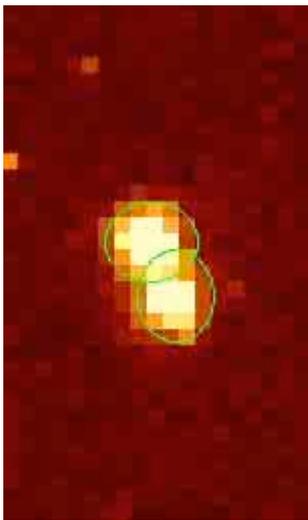

Figure 9d: Sources corresponding to points γ (lower right) and δ (upper left), in 9a as they appear in the raw data.

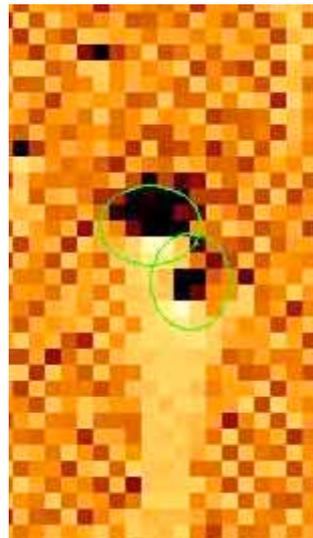

Figure 9e: Sources corresponding to points γ and δ, in 9a as they appear in the difference image generated by the simulation.

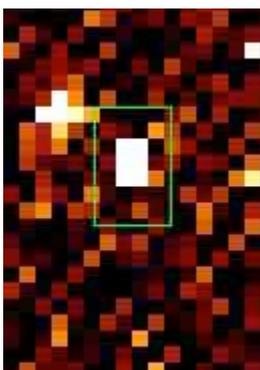

Figure 9f: Source corresponding to point ε, in 9a as it appears in the raw data.

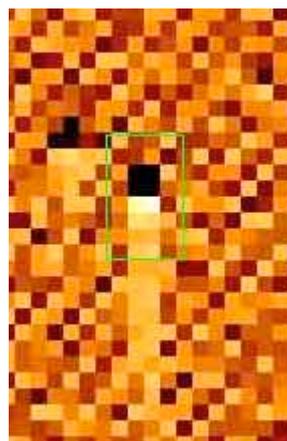

Figure 9g: Source

The source labelled γ is a similar example, it lies below and extremely close to source δ, as can be seen in figure 9d. The smaller simulation derived correction for source γ can then be explained as sources α and β above (see also the difference image, figure 9e). Also apparent here is that δ has a simulation derived correction which is larger than the empirical value. This could be due to the uncertainty in the flux estimate for each source arising from the fact they have been de-blended by the detection software. If δ was assigned a flux which was too great, then the empirical CTE correction (as a fraction of the total flux) would be too low. What is interesting is that figure 9a suggests that if the two objects had not been de-blended, and a single large object had been detected instead, then the empirical and simulation derived corrections would have agreed.

Finally source ε has a simulation derived correction which is somewhat larger than that which the empirical algorithm would assign to it. In figure 9f we see that this source has a far from stellar profile. Instead it has a very sharply defined upper edge. This kind of profile will suffer CTI more acutely than stars (see e.g. Bristow and Alexov 2002), and therefore more than an algorithm which only takes total flux into account would predict (see the considerable trail in figure 9g).

## 5. Summary

In the process of testing our model, we enhanced the pipeline scripts to perform an automatic application of GK2002's empirical corrections to sources detected by SExtractor.

Our CCD readout model is able to correct STIS CCD photometric data and produce results consistent with those obtained with the empirically derived corrections of GK2002 for a range of background and signal levels. The level of CTE trail cleaning can be adjusted by fine tuning some of the model parameters (in particular the trap density normalisation). Visual inspection would suggest that we should increase this slightly. However, the model calibration used was that which is best able to reproduce the empirical correction for photometric sources within a 7 pixel diameter aperture (discussed in this paper) and that for extracted spectra (discussed in Bristow 2003b) as we believe that these relationships are well calibrated and represent the mean for an extensive data set.

The agreement between the simulation derived corrections and empirical corrections plotted in figure 2 could also be easily adjusted by tweaking the simulation parameters. Weakening the model derived corrections would maybe give a better agreement in figure 2, however, as we show with figure 3, the model parameters used give an increasingly accurate fit as we restrict the sample further to bright, stellar like objects. Moreover, moving the parameter normalisation in this direction would damage the agreement to the empirical spectroscopic CTE corrections (Bristow 2003b).

Very often there is a very good reason for sources which do not lie close to the line in figure 2 which can be easily understood by examining the charge distribution in the raw image array as demonstrated in section 4. This is however not always the case, sometimes the reasons are more subtle. We feel confident that the model is simulating the readout of even these objects accurately because of the good match to the empirical corrections on the one hand, and because of the many hat can be clearly understood on the other.

Whilst it would be nice to demonstrate directly agreement between corrected sources, extracted from the bottom of a D amplifier exposure, and their equivalent, extracted from the (almost) CTI free B amplifier exposure, shot noise differences between the two exposures precludes this.

- # References